\documentstyle[prb,aps,multicol,epsfig]{revtex}
\renewcommand{\narrowtext}{\begin{multicols}{2}
\global\columnwidth20.5pc}
\renewcommand{\widetext}{\end{multicols}
\global\columnwidth42.5pc} \multicolsep = 8pt plus 4pt minus 3pt

\newcommand{\twobeone}{\widetext\vskip.6pc \noindent \vrule
width3.375in height.2pt depth.2pt \vrule depth0em height1em\hfill
\vskip.6pc }
\newcommand{\onebetwo}{
\vskip.6pc \indent
 \hfill\vrule depth1em height0pt \vrule width3.375in height.2pt depth.2pt
\vskip.6pc \narrowtext \noindent}

\draft

\title{Azbel-Hofstadter model on triangular lattice revisited}

\author{Min-Fong Yang\cite{email}}

\address{Department of Physics, Tunghai University, Taichung, Taiwan}

\date{\today}

\begin{document}

\maketitle

\begin{abstract}
In the present paper, the mean of Lyapunov exponents for the
Azbel-Hofstadter model on the triangular lattice is calculated. It
is recently proposed that [Phys. Rev. Lett. {\bf 85}, 4920
(2000)], for the case of the square lattice, this quantity can be
related to the logarithm of the partition function of the two
dimensional Ising model and has a connection to the asymptotic
bandwidth. We find that the correspondence of this quantity to the
logarithm of the partition function of the two dimensional Ising
model is not complete for the triangular lattice. Moreover, the
detailed connection between this quantity and the asymptotic
bandwidth is not valid. Thus the conclusions for the mean of
Lyapunov exponents suggested previously depend on the lattice
geometry.
\end{abstract}


\narrowtext

The problem of two-dimensional (2D) Bloch electrons in a
perpendicular magnetic field has long been a fascinating
subject.\cite{Azbel,Hof,Wannier,Wilkinson,Sokoloff,Tcmp,L,WZ,FK,CN,HKW,ATW,Krasovsky}
Besides exhibiting extremely rich energy band structures, it can be
related to various phenomena such as the quantum Hall effect,\cite{a}
the flux-state model for high-$T_{c}$ superconductivity,\cite{b} and
the mean-field transition temperature of superconducting networks or
Josephson junction arrays.\cite{c} In addition to the theoretical
work on this subject, due to the rapid development of nanofabrication
techniques, the experimental indications of the peculiar band
structure have also been reported.\cite{exp1,exp2,exp3,exp_1,exp_2}

The simplest model for this investigation is commonly referred to as
the Azbel-Hofstadter model.\cite{Azbel,Hof} Although much work has
been carried out, the analytic description of the spectrum remains
open to a large extent. Recently, some progresses are made to relate
this model to the integrable systems, and thus deepen our
understanding.

It was noticed that the model can be related to integrable systems
with a quantum group algebra.\cite{WZ,FK} Hence explicit solutions
to this problem can be constructed.\cite{HKW,ATW} Recently,
Krasovsky shows that a particular quantity, the mean of Lyapunov
exponents, plays an important role in the study of the
Azbel-Hofstadter problem and gives a different type of relation to
integrable systems.\cite{Krasovsky} It is found that, for any
value of magnetic flux $\Phi=2\pi P/Q$ through a plaquette ($P$
and $Q$ are mutually prime integers), this quantity can be related
to the logarithm of the partition function of the 2D Ising model.
Moreover, the band edges of the Azbel-Hofstadter Hamiltonian
correspond to the critical temperature of the Ising model. Besides
the relation to the solvable 2D Ising model, it is proposed that
the mean of Lyapunov exponents can also be connected to the
asymptotic (large $Q$) bandwidth. These results provide a very
exciting connection between two important models, and will create
a perspective for even deeper understanding, if they are universal
irrespective of various kinds of lattice structures. For example,
it is well-known that the concept of universality was discovered
through the fact that the critical exponents of phase transitions
in the Ising model are independent of lattice geometry. Thus it is
interesting to investigate if these conclusions are still valid
for other kinds of 2D lattice structures.

The aforementioned theoretical and experimental studies focused on
the energy spectrum of a lattice with a square symmetry. Triggered
by fascinating results for the square lattice, other work has
considered various kinds of 2D lattices in order to investigate
the effect of the lattice geometry on the energy spectrum. Besides
the square lattice, the most extensively studied case has been the
triangular
lattice.\cite{Langbein,CW,Claro,T83,HHKM,HK,WA,BKS,Han_et_al,GF,Oh}
The energy spectrum of the triangular lattice with isotropic
hopping is known to exhibit a recursive band structure as in the
case of the square lattice.\cite{CW} Moreover, it has been shown
that,\cite{T83,Han_et_al} the same as it is for the square
lattice, the sum of all bandwidths $W\sim 32G/\pi Q$ as
$Q\to\infty$, where $G = 0.91596559417721...$ is Catalan's
constant. (Notation $A\sim B$ means, henceforth, that $A/B$ tends
to 1 in the limit.) However, in contrast to the case of the square
lattice, it is shown that, for the case of the triangular lattice,
not only the symmetry of the energy spectrum is lowered, but also
a degeneracy at the band center of the energy spectrum is
removed.\cite{Claro,HK,Han_et_al}

Motivated by Krasovsky's work,\cite{Krasovsky} in the present
paper, we calculate the mean of Lyapunov exponents for the case of
the triangular lattice, and investigate if Krasovsky's conclusions
will depend on lattice structures. For simplicity, we assume
isotropic hopping and set the hopping amplitude to be unity. We
find that the mean of Lyapunov exponents of the Azbel-Hofstadter
model still has a similar form to the logarithm of the partition
function of the 2D Ising model. However, in contrary to the case
of the square lattice, we show that the band edges of the
Azbel-Hofstadter model do not always correspond to the critical
temperature of the Ising model in the present case. It means that
the identification between these two models is not complete in the
case of the triangular lattice. Later, another integral form of
the mean of Lyapunov exponents is derived. From this integral
representation, it is easily found that the mean of Lyapunov
exponents at zero energy is positive, rather than zero as in the
case of the square lattice. From this observation, we conclude
that, while this quantity and the asymptotic bandwidth both scale
as $1/Q$, a detailed relation between these two quantities, which
is conjectured by Krasovsky, fails in the case of the triangular
lattice. Thus the aforementioned role played by the mean of
Lyapunov exponents in the study of the Azbel-Hofstadter problem
depends on the the lattice geometry.

In the investigation of the Azbel-Hofstadter problem, there are
two complementary, but mathematically equivalent, approaches: one
can either consider the strong magnetic field limit with the
lattice potential treated as a small perturbation,\cite{Azbel} or
study the influence of a weak magnetic field on the strong lattice
potential in the tight-binding approximation.\cite{Hof} Here we
take the latter approach. For the tight-binding case, the model on
the triangular lattice is topologically equivalent to that on the
$N\times N$ square lattice with the next-nearest-neighbor hopping
in only one direction.\cite{T83,HHKM,HK,Han_et_al} When the
magnetic flux through a plaquette is $2\pi P/Q$, the quasi-momenta
$k_x^0=2\pi m/N$ and $k_y=2\pi k/N$ are defined in the magnetic
Brillouin zone: $0 \leq k_x^0 <2\pi /Q$ and $0 \leq k_y < 2\pi$
(that is, $m=0,1,\dots, N/Q-1$ and $k=0,1,\dots,N-1$). The matrix
of the Azbel-Hofstadter Hamiltonian $H$ can be decomposed into a
direct sum of the $Q\times Q$ matrices $H_{km}$ for each $k$ and
$m$. In the case of the triangular lattice, the dependence of the
spectral determinant (characteristic polynomial) of $H_{km}$ on
$k$ and $m$ can be shown as (for example, see Eq.~(4.1) in
Ref.~\onlinecite{HK})
\twobeone
\begin{equation}\label{ch}
\left|\det(H_{km}-\varepsilon I_{Q\times Q})\right|=
\left|\left[\sigma(\varepsilon)-2\cos(\frac{2\pi
m}{N/Q})-2\cos(\frac{2\pi k}{N/Q}) + 2 (-1)^{Q+P} \cos(\frac{2\pi
k}{N/Q}+\frac{2\pi m}{N/Q}) \right]\right|,
\end{equation}
where $\sigma(\varepsilon)=\varepsilon^Q + \dots$ is a polynomial
of $Q$th degree whose coefficients do not depend on $k$ and $m$.
Therefore,
\begin{eqnarray}\label{det}
&&\left| \det(H-\varepsilon I) \right|
=\prod_{k=0}^{N-1}\prod_{m=0}^{N/Q-1} \left| \det(H_{km}-\varepsilon
I_{Q\times Q}) \right| \nonumber \\ &&=\prod_{k,m=0}^{N/Q-1} \left|
\sigma(\varepsilon)-2\cos(\frac{2\pi m}{N/Q})-2\cos(\frac{2\pi
k}{N/Q}) + 2 (-1)^{Q+P} \cos(\frac{2\pi k}{N/Q}+\frac{2\pi m}{N/Q})
\right|^Q.
\end{eqnarray}

In the limit of infinite lattice ($N\to\infty$), one can replace
$2\pi m Q/N$ and $2\pi k Q/N$ by continuous parameters $x$ and
$y$, respectively. As discussed in Ref.~\onlinecite{Krasovsky},
the mean of Lyapunov exponents $\gamma_-$ for the case of $Q+P=$
odd integers can be written as
\begin{eqnarray}\label{pf1}
\gamma_-(\sigma) &=& \lim_{N\to\infty}{1\over
N^2}\ln|\det(H-\varepsilon I)| \nonumber \\ &=&{1\over 4\pi^2
Q}\int_0^{2\pi} dx \int_0^{2\pi} dy \ln \left|
\sigma(\varepsilon)- 2 \left[ \cos x + \cos y + \cos(x+y) \right]
\right|.
\end{eqnarray}
For the case of $Q+P=$ even integers, the mean of Lyapunov
exponents $\gamma_+$ can be defined in the same way. By using the
periodicity of the cosine function, one can show that
$\gamma_\pm(-|\sigma|) = \gamma_\mp(|\sigma|).$ Thus we need only
to consider the case of $\gamma_-$ (i.e., the case of $Q+P=$ odd
integers). We note that the energy spectrum for $Q+P=$ odd
integers is determined by the secular equation,
$\sigma(\varepsilon) = 2 \left[ \cos x + \cos y + \cos(x+y)
\right]$. Because $-3 \leq 2 \left[ \cos x + \cos y + \cos(x+y)
\right] \leq 6$, following the same discussion in
Ref.~\onlinecite{Krasovsky}, one can show that the spectrum of $H$
in the limit $N\to\infty$ consists of $Q$ bands, which are just
the image of the interval $[-3, 6]$ under the inverse of the
transform $\sigma=\sigma(\varepsilon)$. In particular, the band
edges $\varepsilon_{e_i}$ satisfy $\sigma(\varepsilon_{e_i})= -3$
or $6$.

On the other hand, for the 2D ferromagnetic Ising model with
isotropic interactions on a triangular lattice, the partition
function $Z$ satisfies:\cite{Ising_reviews}
\begin{equation}\label{pf2}
\lim_{N\to\infty}{1\over N^2}\ln Z-\ln 2 ={1\over
8\pi^2}\int_0^{2\pi} dx \int_0^{2\pi} dy  \ln \left| C^3 + S^3 - S
\left[ \cos x + \cos y + \cos(x+y) \right] \right|.
\end{equation}
\onebetwo
Here $C=\cosh(2J/T)$ and $S=\sinh(2J/T)$, where $T$ is
the temperature and $J$ is the interaction constant. It is found
that Eqs.~(\ref{pf1}) and (\ref{pf2}) have similar integral
representations. Thus one may conclude that the connection between
these two models still holds in the case of the triangular
lattice. Nevertheless, in contrast to the case of square lattice,
we find that the band edges of $H$ do not always correspond to the
critical temperature of the Ising model for the triangular
lattice. We note that, because $(C^3 + S^3) / S \geq 3$, the
integrand in Eq.~(\ref{pf2}) can be divergent only at the critical
temperature such that $C^3 + S^3 =3S$, and it diverges at the
upper/lower limits of integration, $x=0$, $2\pi$ and $y=0$,
$2\pi$. For the Azbel-Hofstadter Hamiltonian, at the band edges
$\varepsilon_{e_i}$ satisfying $\sigma(\varepsilon_{e_i})= 6$, the
integrand in Eq.~(\ref{pf1}) also diverges at $x=0$, $2\pi$ and
$y=0$, $2\pi$. Thus there is a correspondence between these band
edges of $H$ and the critical temperature of the Ising model.
However, at the band edges $\varepsilon_{e_i}$ satisfying
$\sigma(\varepsilon_{e_i})= -3$, the integrand in Eq.~(\ref{pf1})
diverges only at $x=y=2\pi /3$ or $4\pi /3$. It means that the
band edges $\varepsilon_{e_i}$ satisfying
$\sigma(\varepsilon_{e_i})= -3$ have no correspondence to the
critical temperature of the Ising model. Therefore, the
identification between these two models on the triangular lattice
is not complete.

It is noted by Krasovsky that,\cite{Krasovsky} the mean of Lyapunov
exponents is proportional to $1/Q$ (see also Eq.~(\ref{pf1}) in the
present paper). Since the total bandwidth is shown to be of order
$1/Q$ as $Q\to\infty$,\cite{L} it is reasonable to think of a
connection between these two quantities. For the case of the square
lattice, Krasovsky shows that the mean of Lyapunov exponents at the
band edges is eight times less than the estimated asymptotic (large
$Q$) bandwidth.\cite{T83,Tcmp} By defining $W(x)$ to be the total
length of the image of $[0,x]$ under the inverse of the transform
$\sigma=\sigma(\varepsilon)$, he further conjectures that
$W(\sigma)\sim 4\gamma(\sigma)$ for any coprime $P$ and $Q$ as
$Q\to\infty$. It is interesting to investigate if this conjecture is
still valid for the triangular lattice.

For further discussions, it is more convenient to recast
Eq.~(\ref{pf1}) into another from. By combining the terms $\cos y$
and $\cos(x+y)$, and then taking the integrals over $y$, we
obtain\cite{note}
\begin{eqnarray}\label{gamma_2}
\gamma_-(\sigma=-2)&=&{1\over 2\pi Q} \int_0^{2\pi} \ln  \left\{ 2
\left| \cos(\frac{x}{2}) \right| \right\} dx \nonumber \\ &=&
{1\over 2\pi Q} \int_0^{\pi} \ln \left( 2 + 2\cos x \right) dx =0.
\end{eqnarray}
Hence for $-2 < \sigma < 6$, one has
\begin{equation}\label{recast1}
\gamma_-(\sigma)=\int_{-2}^{\sigma}  \frac{d \gamma_-(x)}{dx}
\;dx,
\end{equation}
where
\begin{eqnarray}\label{d_gamma1}
&&\frac{d \gamma_-(\sigma)}{d\sigma} \nonumber
\\ &=&\frac{1}{2\pi Q}
\int_{\cos^{-1}(1-\sqrt{\sigma +3}+ {\sigma \over 2})}^\pi
\frac{dx}{\sqrt{(\cos x -{\sigma \over 2})^2 - 4\cos^2(\frac{x}{2})}}
\nonumber
\\&=&\frac{1}{\pi Q} \frac{ K \left[ \frac{(\nu-1)
\sqrt{(\nu-1)(\nu+3)}}{4\sqrt{\nu}} \right]}{2\sqrt{\nu}} \; .
\end{eqnarray}
Here $\nu \equiv \sqrt{\sigma+3}$, and
$K(k)=\int_0^{\pi/2}(1-k^2\sin^2 x)^{-1/2}dx$ is the complete
elliptic integral of the first kind. The first equation in
Eq.~(\ref{d_gamma1}) is derived by employing the standard result
\begin{equation}
\frac{1}{2\pi} \int_0^{2\pi} \frac{d\phi}{\zeta - \cos \phi} =
\cases{ (\zeta^2 -1)^{-1/2}, &\text{$\zeta > 1$} ;\cr -(\zeta^2
-1)^{-1/2}, &\text{$\zeta < -1$};\cr 0, &\text{$|\zeta|\le 1$},}
\end{equation}
and the last equation in Eq.~(\ref{d_gamma1}) is obtained by using
the formula\cite{Bateman}
\begin{eqnarray}
&&\int_\delta^\gamma
\frac{dx}{\sqrt{(\alpha-x)(\beta-x)(\gamma-x)(x-\delta)}} \nonumber
\\ &=& \frac{2 K \left[
\sqrt{
\frac{(\alpha-\beta)(\gamma-\delta)}{(\alpha-\gamma)(\beta-\delta)} }
\right]}{\sqrt{(\alpha-\gamma)(\beta-\delta)}} \; .
\end{eqnarray}
 Similarly, for $-3 < \sigma <
-2$, we obtain
\begin{equation}\label{recast2}
\gamma_-(\sigma)= - \int_{\sigma}^{-2}  \frac{d \gamma_-(x)}{dx}
\;dx,
\end{equation}
where
\begin{eqnarray}\label{d_gamma2}
&&\frac{d \gamma_-(\sigma)}{d\sigma} \nonumber
\\&=&-\frac{1}{2\pi Q}
\int_{\cos^{-1}(1-\sqrt{\sigma +3}+ {\sigma \over 2})}^\pi
\frac{dx}{\sqrt{(\cos x -{\sigma \over 2})^2 - 4\cos^2(\frac{x}{2})}}
\nonumber
\\ &&- \frac{1}{2\pi Q} \int_0^{\cos^{-1}(1+\sqrt{\sigma +3}+ {\sigma \over 2})}
\frac{dx}{\sqrt{(\cos x -{\sigma \over 2})^2 - 4\cos^2(\frac{x}{2})}}
 \nonumber
\\&=&-\frac{4}{\pi Q} \frac{K \left[ \frac{(1-\nu)
\sqrt{(1-\nu)(3+\nu)}}{(1+\nu) \sqrt{(1+\nu)(3-\nu)}}
\right]}{(1+\nu) \sqrt{(1+\nu)(3-\nu)}} \;.
\end{eqnarray}
From these formulas, the mean of Lyapunov exponents at the band
edges can be numerically calculated with the results:
$\gamma_-(\sigma=6) \cong 5.075/\pi Q$ and $\gamma_-(\sigma=-3)
\cong 2.030/\pi Q$. Thus $4 \gamma_-(\sigma=6) + 4
\gamma_-(\sigma=-3)\cong 28.42 /\pi Q$, which is close to the
estimated value of the total width $W = W(-3)+W(6) \sim 32G/\pi Q
\cong 29.31 /\pi Q$ as $Q\to\infty$.\cite{T83,Han_et_al} This
seems to support the validity of Krasovsky's conclusion in the
case of the triangular lattice. Nevertheless, because
$d\gamma_-(\sigma) / d\sigma \geq 0$ for $-2 < \sigma < 6$, we
find from Eq.~(\ref{recast1}) that $\gamma_-(\sigma=0)$ is
positive rather than zero. (One can show numerically that
$\gamma_-(\sigma=0) \cong 1.347/\pi Q$.) On the other hand, it is
obvious that $W(0)$, which is the total length of the image of a
zero interval under the inverse of the transform
$\sigma=\sigma(\varepsilon)$, must be zero. Therefore, the
detailed relation, $W(\sigma)\sim 4\gamma(\sigma)$, which is
conjectured in Ref.~\onlinecite{Krasovsky}, fails in the present
case.

In summary, the integral expressions of the mean of Lyapunov
exponents for the case of the triangular lattice is obtained in
the present work. From these expressions, we show that it can not
be properly related to the logarithm of the partition function of
the 2D Ising model. Moreover, the conjectured relation
$W(\sigma)\sim 4\gamma(\sigma)$ can not be true for the general
types of lattice structures. Therefore, the mean of Lyapunov
exponents plays a less important role in the study of the
Azbel-Hofstadter problem on the triangular lattice.

\acknowledgments I am grateful to M.~C.~Chang for encouraging my
interest in the problem and for valuable comments. I also thank
I.~V.~Krasovsky for the discussions of his paper. Finally, I
acknowledges financial support by the National Science Council of
Taiwan under Contract No. NSC 89-2112-M-029-006.

\widetext


\begin{references}
\bibitem{note}
Eq.~(\protect\ref{gamma_2}) is obtained by using the following
formula: $I = \int_0^{2\pi} \ln |a+b\cos y| dy = 2\pi\ln \left[
\left(|a|+\sqrt{a^2-b^2} \right)/2 \right]$, if $|a| \geq |b|$; and
$I = 2\pi\ln \left(|b|/2 \right)$, if $|a|<|b|$. This formula can be
derived by differentiating with respect to $b$ and then doing the
integration in the complex plane. I learns this formula from Dr.
I.~V.~Krasovsky.
\end{references}
\end{document}